\documentclass[twocolumn, 11pt]{article}
\usepackage[text={7.2in,8.5in},centering]{geometry}
\usepackage{hyperref}
\hypersetup{
     colorlinks = true,
     linkcolor = linkcolor,
     anchorcolor = linkcolor,
     citecolor = linkcolor,
     filecolor = linkcolor,
     urlcolor = linkcolor  
     }

\usepackage{physics}
\usepackage{amsfonts, amsmath, amssymb}
\usepackage{float}
\usepackage{acronym} 
\usepackage{cuted}
\usepackage{times}
\usepackage{graphicx}
\usepackage{latexsym}
\usepackage{url}
\usepackage{listings}
\usepackage[labelfont=bf,font={small}]{caption}
\usepackage{comment}
\usepackage{xcolor}
\usepackage{amsmath}
\usepackage{tabularx}
\usepackage{adjustbox}
\usepackage{framed}
\usepackage{comment}
\usepackage{colortbl}
\usepackage{setspace}
\usepackage{subcaption}
\usepackage{booktabs}
\usepackage{makecell}
\usepackage{siunitx}
\usepackage{parskip}
\usepackage{tcolorbox}
\usepackage{lscape}
\usepackage{enumitem}
\usepackage{makecell}
\usepackage{graphicx}
\usepackage{latexsym}
\usepackage{lipsum}
\usepackage{graphicx}
\definecolor{linkcolor}{RGB}{0,50,200}

\graphicspath{{Figures/}}

\usepackage{fancyhdr}
\makeatletter
\AtBeginDocument{%
  \renewcommand*{\AC@hyperlink}[2]{%
    \begingroup
      \hypersetup{hidelinks}%
      \hyperlink{#1}{#2}%
    \endgroup
  }%
}
\makeatother

\title{\vspace*{-1cm}Optical Convolutional Neural Networks -- Combining Silicon Photonics and Fourier Optics for Computer Vision} 
\date{}

\author{
	Cottle, Edward\\
	\texttt{edward.cottle@optalysys.com}
	\and
    Michel, Florent\\
	\texttt{florent.michel@optalysys.com}
	\and
	Wilson, Joseph\\
	\texttt{joseph.wilson@optalysys.com}
	\and
	New, Nick\\
	\texttt{nick.new@optalysys.com}
	\and
	Kundu, Iman\\
	\texttt{iman.kundu@optalysys.com}
    \\ \\
    \small{\textit{Optalysys Ltd\footnote{\url{www.optalysys.com}}, 8 Flemming Court, Wakefield WF10 5HW, United Kingdom}}
}
	
\begin{document}

	\thispagestyle{fancy}
	\maketitle
	\begin{strip}
	\vspace{-12ex}
    \begin{abstract}
	    The Convolutional Neural Network (CNN) is a state-of-the-art architecture for a wide range of deep learning problems, the quintessential example of which is computer vision. CNNs principally employ the convolution operation, which can be accelerated using the Fourier transform. In this paper, we present an optical hardware accelerator that combines silicon photonics and free-space optics, leveraging the use of the optical Fourier transform within several CNN architectures. The hardware presented is a proof of concept, demonstrating that this technology can be applied to artificial intelligence problems with a large efficiency boost with respect to canonical methods.
    \end{abstract}
    \vspace{6ex}
    \end{strip}
  
    \section{Introduction}
    	
    The acceleration of convolutions and correlations using optical processing is a well established technique. In 1964 Vander Lugt demonstrated the application of complex filters to the problem of pattern matching using a 4f optical correlator~\cite{lugt1964signal}. Similarly, the joint-transform correlator was first presented by Weaver and Goodman in 1966, who outlined an optical means of correlating two signals with a 2f system~\cite{weaver1966technique}. These approaches rely on the property that an ordinary lens produces (in the Fresnel approximation of wave optics) the \ac{FT} of the complex amplitude of an incident electro-magnetic field, which we refer to as the \ac{OFT}. 
    \par
    Optalysys is currently developing a hybrid \ac{SiPh} driven free-space optical processing device that also exploits this property. The device demonstrates optical encoding of digital data using a thermally modulated \ac{MZI} array using \ac{SiPh}. These interferometers control the phase and amplitude of light, allowing complex data to be encoded in a 2-dimensional grid, the \ac{OFT} of which is subsequently obtained using a free-space optical stage. This approach yields a dynamically configurable device in a micro-scale form factor suitable for integration with digital electronics.
    This demonstrator device has been used to implement the first optical \ac{CNN}~\cite{lecun1995convolutional} leveraging free-space optics and \ac{SiPh} as an alternative to digital processing. The primary aim of this paper is to report the results we have obtained and outline the significant advantages that our optical approach offers the field of AI. \par
    
    The \ac{FT} is commonly used for digital signal processing, as it converts a signal into its frequency domain representation. The convolution of a signal can be achieved by  an efficient point-wise multiplication in the frequency domain that scales linearly, a significant improvement over the quadratic algorithmic complexity of the naive implementation in the spatial/temporal domain~\cite{mcgillem1991continuous}. Thus, to improve on efficiency, convolution operations are often carried out using the \ac{FT}. \par
    
    While a proof of the lower bounds on the algorithmic complexity of the \ac{FT} is still an open problem in computer science, the most efficient algorithm to date, the \ac{FFT}~\cite{cooley1965algorithm}, has complexity $\mathcal{O}(n \log n)$ where $n$ is the number of discrete data elements to be transformed. This algorithmic complexity is the same for one and two-dimensional signals, and provides a reduction from the quadratic algorithmic complexity for convolution operations, justifying wide-spread adoption in modern electronic implementations. In contrast, passing an optical signal through a lens has a notional algorithmic complexity of $\mathcal{O}(1)$; the processing time required to obtain the \ac{FT} of the signal is the same, no matter how much information is encoded into the light. As the calculation is performed through light propagation via interference, the processing time is simply the time taken for the light to complete a transit of the free-space optical system. Using an \ac{OFT} can therefore further reduce the algorithmic complexity to a linear relationship, in which the point-wise multiplications can be carried out in parallel. Therefore, provided a processor that can fully parallelise these multiplications, the entire convolution operation can be carried out in a single clock cycle of an optical system with a runtime complexity of $\mathcal{O}(1)$. \par
        
    Applied as a co-processor in a digital system, the capabilities of an optical calculation method are therefore bounded only by the speed at which information can be transferred to and from the optical regime. \ac{SiPh} is a well established technology in the field of communications and integrated \ac{SiPh} components are capable of exceptionally high data rates. As applied to the problem of Fourier-optical processing, a \ac{SiPh} approach yields throughput that vastly exceeds the most efficient electronic equivalents. \par
    
    The input size and frame rate achievable with the current demonstrator device are not representative of what ongoing and future developments will achieve. For this reason, we do not report training and inference time, but focus instead on the accuracy reached by the network. Indeed, the question of whether free-space optical computing, with its inherent sensitivity to external noise and fluctuations, can be used in practice to implement deep neural networks is, to the best of our knowledge, still open. As this work will show, building and training a neural network on such an optical device is indeed possible, and can achieve results rivaling electronic implementations. \par
        
    \subsection{Fourier optics}
        
    The \ac{OFT} holds in the Fresnel approximations of light diffraction. We begin by choosing a Cartesian coordinate system $(x,y,z)$ and assuming there is no incoming light from the region $z \to \infty$. By applying the Huygens principle~\cite{huygens,crew2009wave} in the plane $z = z_1 < z_2$, the complex amplitude of a monochromatic electro-magnetic wave in the plane $z = z_2$ can be obtained by evaluating the following integral (considering for simplicity that there is only one polarization mode):
    \begin{equation}
    \begin{aligned}
    	\psi(x, y, z_2) \propto \iint_{x',y'=-\infty}^{+\infty} & K(x-x', y-y',z_2-z_1) \\ 
    	& \psi(x', y', z_1) \, \mathrm{d} x' \mathrm{d} y',
    	\label{eq:polarmode}
    \end{aligned}
	\end{equation}
    where $K$ is a function from $\mathbb{R}^3$ to $\mathbb{R}$ defined by:
    \begin{equation}
        K(x, y,z) =
        \frac{\exp \left[ \frac{i n \omega}{c} \sqrt{x^2 + y^2 + z^2} \right]}{\sqrt{x^2 + y^2 + z^2}},
        \label{eq:k_function}
    \end{equation}
    Here, $n$ is the optical index of the medium, assumed to be uniform, $\omega$ is the angular frequency of the wave, $c$ is the speed of light in vacuum, and $i$ a square root of $-1$. 
    The symbol $\propto$ indicates that this expression is correct up to a global normalization factor, which we omit for simplicity. A thin lens placed in the plane $z=0$ will add a dephasing proportional to the lens width and variation of optical index it induces. If the surface of the lens is parabolic close to $x = y = 0$, the integral over the coordinates in this plane is Gaussian and yields, after a few lines of algebra:
    \begin{equation} \label{eq:apprOFT}
    \begin{aligned}
       \psi_f(x, y) \propto \mathcal{F}(\psi_{-f}) \left( \frac{n \omega}{c f} x,  \frac{n \omega}{c f} y\right) \\
        \left( 
            1
            + O \left( \frac{x^2}{f^2} \right)
            + O \left( \frac{y^2}{f^2} \right)
        \right),
    \end{aligned}
    \end{equation}
    where $f$ is called the focal length of the lens, $\mathcal{F}$ denotes the \ac{FT} in the directions $x$ and $y$, and we denote by $\psi_z$ the complex amplitude restricted to a plane of fixed coordinate $z$. 
    We refer the interested reader to the textbooks~\cite{Goodman,ghatak2005optics} for details of the derivation.
    	
    In short, Equation~\eqref{eq:apprOFT} means that the amplitude of the field in the plane $z=f$ is the \ac{FT} of that in the plane $z = -f$. 
    While derived here for a stationary wave, in practice Equation~\eqref{eq:apprOFT} is a very good approximation in the time-dependent case too, provided that the input field $\psi_{-f}$ changes on a time scale that is long before the time $\tau = 2 f n / c$. 
    Taking $n$ close to $1$, $f$ of a few centimeters, and using $c = 299{,}792{,}458\mathrm{m/s}$, we obtain $\tau \approx 10^{-8} \mathrm{s}$. 
    	
    We emphasise that this is only a limit on the latency of the  \ac{OFT}, not on its throughput. The latter is only limited by the frequency of the wave and time delay between different light rays. Using light in the near infrared or visible range and for values of $x$ and $y$ of the order of a millimeter, the theoretical `frame rate' limit of such a system is well above 100 GHz. Combined with the fully parallel nature of the \ac{OFT}, and considering an optical device with a square input grid with tens of pixels per side, the theoretical bound on the throughput can reach hundreds of terabits per second. We will not focus on this point in the present article, as the current device was not designed for optimal speed. However, the high theoretical throughput and low power consumption compared with electronic hardware are important motivations for optical computing in general, and our study in particular. 
    	
    \subsection{Silicon Photonics as a Display Device}
    
    We have developed prior systems that also compute optical convolutions. These systems made use of \ac{SLM} \ac{LCD} devices. The \ac{SLM} \ac{LCD} devices provide a two-dimensional grid of pixels which can achieve controllable absorption and phase changes, providing a basis for optical multiplication. When paired with a 4f correlator architecture, such devices can be used to perform the convolution of two signals in a single optical frame. While capable of the simultaneous processing of several million data-elements, the main drawbacks of these \ac{SLM} devices are their limited phase precision and their limited bandwidth (typically on the order of Hz or kHz - due to the phase change rise-and-fall time characteristics of the liquid crystal medium). Furthermore, the feature-extraction kernels applied in \ac{CNN} architectures typically span only 3-7 data elements in both dimensions, reducing the advantage of using large grids for deep learning.
    	
    Given the two main limiting factors on the speed of an optical \ac{FT} accelerator (the bandwidth of light and the encoding/decoding speed), we have created a new kind of display device that can push these limitations as far as is feasible with current technology. \ac{SiPh} provides a mechanism by which data can be encoded at rates exceeding 50 Gbps with conventional components. Until now the combination of \ac{SiPh} and free-space optics has not been employed. However, as we will show, this has now been realised by the creation of our first prototype system.
    	
    \begin{figure}
        \includegraphics[width=0.5\textwidth]{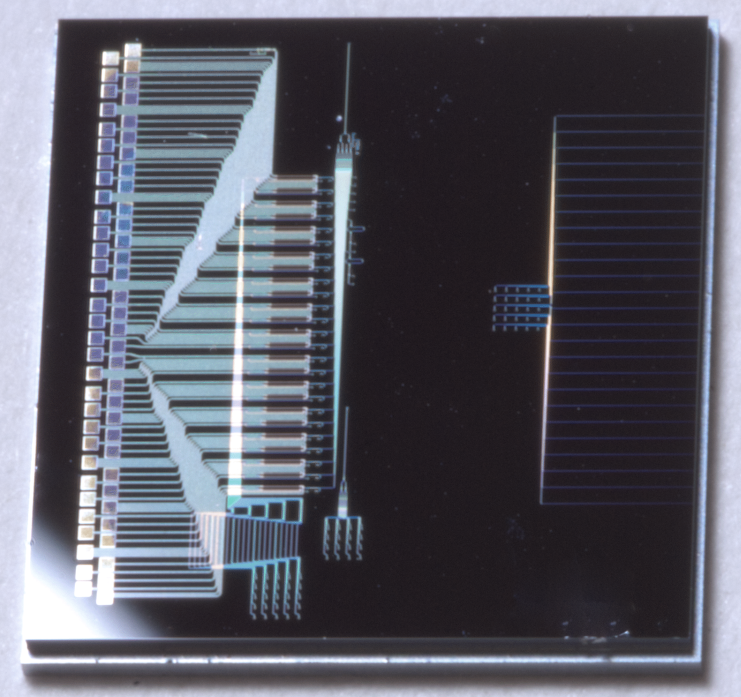} 
        \caption{Image of the \ac{SiPh} section for encoding complex digital information into light.}
        \label{fig:siph}
    \end{figure}
    	
    In our system, data is encoded into low power infra-red laser light in parallel \ac{SiPh} waveguides via thermally-modulated \ac{MZI} units. The waveguides are then arranged into a two-dimensional grid wherein light is emitted from a set of grating couplers into a free-space optical section, where a lens produces the \ac{FT} of the encoded signal. The light is then coupled back into a waveguide array using a two-dimensional grid of grating couplers (see Figure~\ref{fig:siph}). Once returned to the \ac{SiPh} medium the optical signal can be detected and converted back into digital data. A complete description of the device and experimental set-up will be given in a future publication.
    
    \begin{figure}
        \includegraphics[width=0.5\textwidth]{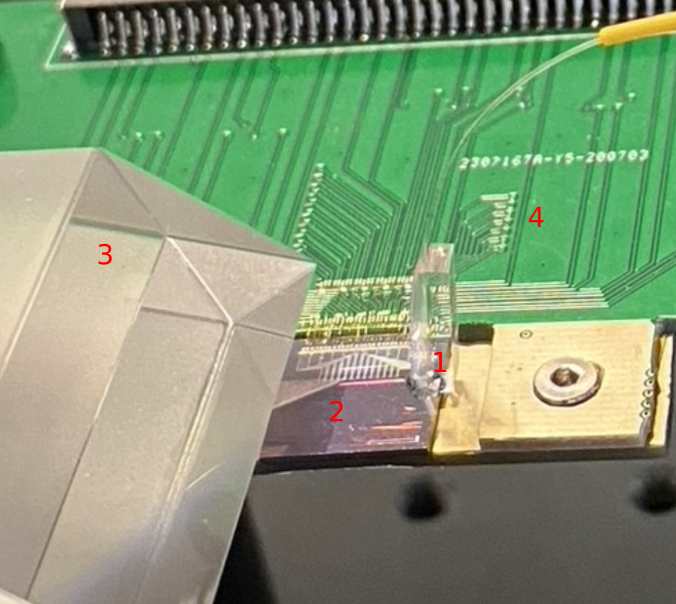} 
        \caption{Image of the experimental setup: 1) Coherent laser source input. 2) Silicon photonics section. 3) Free-space optics section. 4) Electronic drive board.}
        \label{fig:experimental_setup}
    \end{figure}
    
    \begin{figure}
        \includegraphics[width=0.5\textwidth]{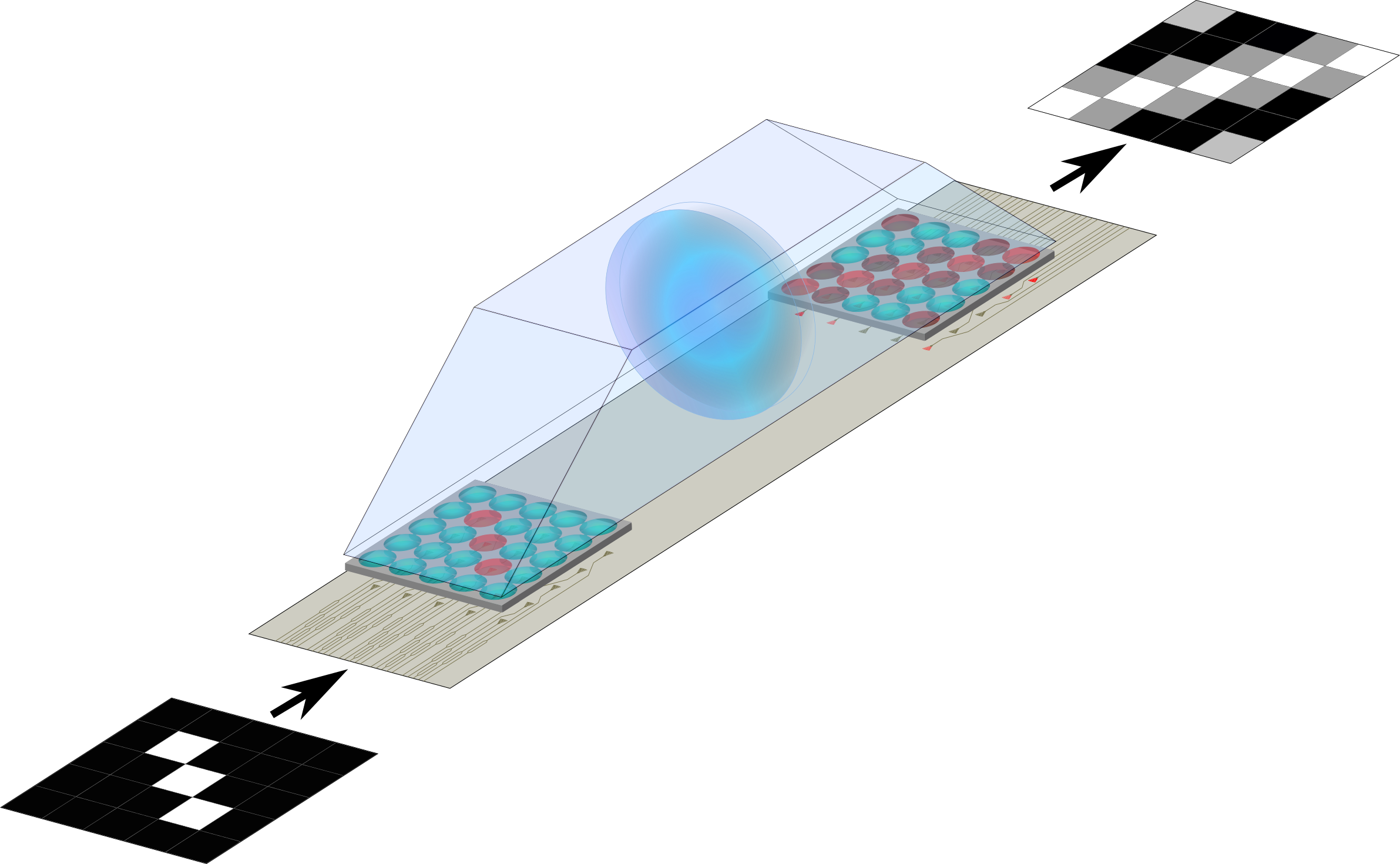} 
        \caption{Conceptual diagram of the optical device. Information is encoded into light via array of Mach-Zehnder interferometers (bottom left). Light passes through lens within solid free-space section (centre). \ac{FT} of input is captured into second waveguide array (top right).}
        \label{fig:system_diagram}
    \end{figure}
    
    \begin{figure}
        \includegraphics[width=0.5\textwidth]{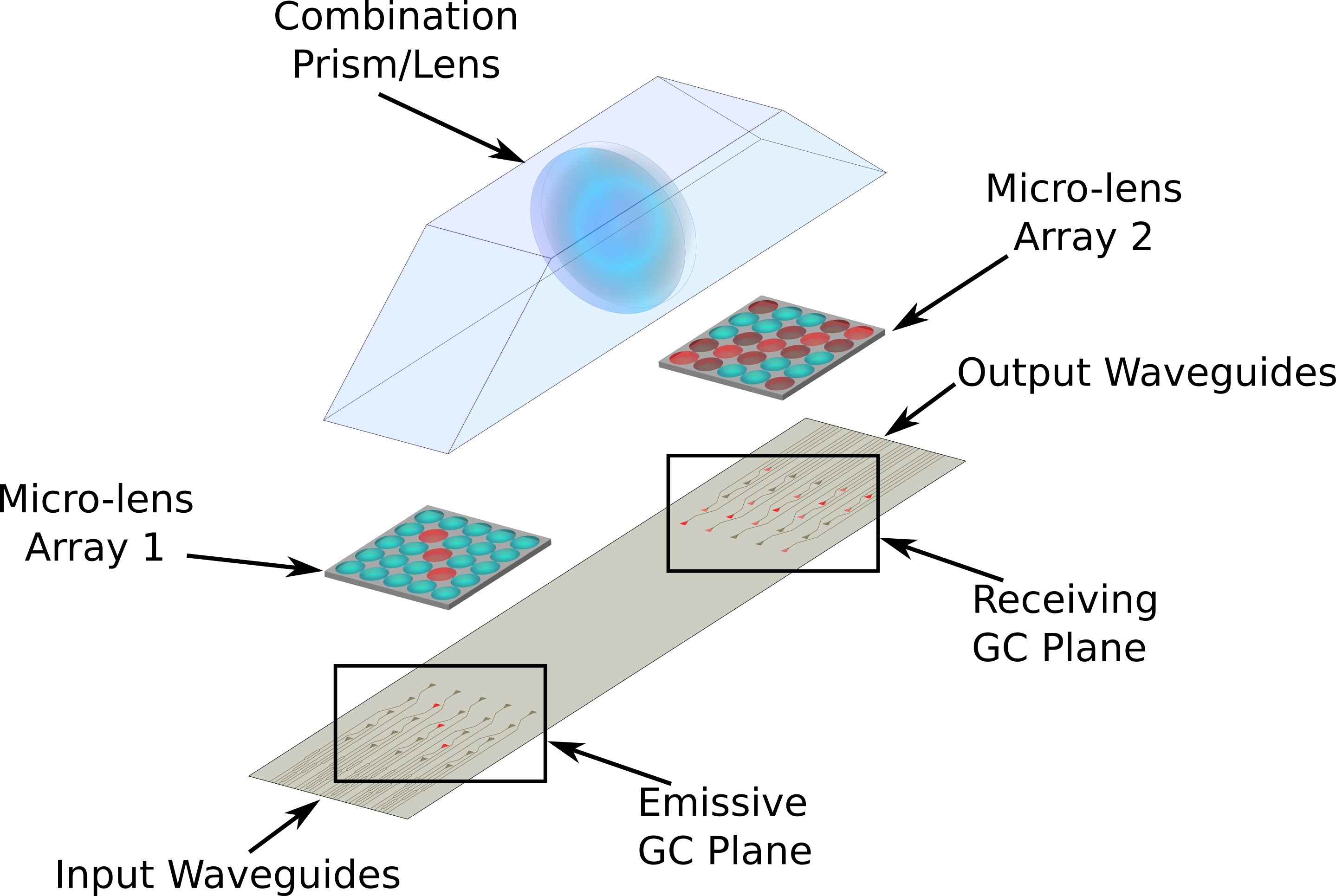} 
        \caption{Exploded diagram of the optical device concept.}
        \label{fig:exploded_system_diagram}
    \end{figure}
    
    The system is designed to perform detection by means of a balanced photodiode array capable of the simultaneous retrieval of both both amplitude and phase information; however, for this study detection was performed using an InGaaS Short-Wavelength Infra-Red (SWIR) camera. As the camera detects the intensity of the optical signal, the output is the absolute value of the square of the amplitude and the phase data is not detected. Despite this we are able to obtain the complex \ac{FT} result using the following identity:
    
    \begin{equation}
        \begin{aligned}
            \mathcal{F}(x) = \frac{1}{4\mathcal{F}(y)^\star}[|\mathcal{F}(x+y)|^2 - |\mathcal{F}(x-y)|^2 + \\ i(|\mathcal{F}(x+iy)|^2 - |\mathcal{F}(x-iy)|^2)],
        \end{aligned}
    \end{equation}
    
    where $x$ and $y$ are two arrays of values. 
        
    To determine the left hand side of the equation, we take 4 passes through the system encoding our desired input $x$, with an arbitrary value $y$ selected. The arrays $y$ and $iy$ are added and subtracted from $x$ as above and the resultant outputs from the machine are combined to produce the full complex \ac{FT} of $x$.  
    	
    \subsection{The Multiply \& Fourier Transform Unit}
    	
    The convolution theorem \eqref{eqn:convolution_theorem} describes how the convolution of two signals may be computed via multiplication in the frequency domain. This property allows the use of an optical \ac{MFT} unit to achieve a convolution between an image and kernel function as in a convolutional layer in a \ac{CNN} \cite{mcgillem1991continuous}. (See equations \eqref{eqn:mft_operation} and \eqref{eqn:mft_operation_to_convolution}.) For all 2D arrays $x$ and $y$, we have the following identity:
	
    \begin{equation}
    \label{eqn:convolution_theorem}
        \begin{aligned}
            x * y = \mathcal{F}^{-1}(\mathcal{F}(x) \cdot \mathcal{F}(y))
        \end{aligned}
    \end{equation}
    
    For all 2D arrays $x$ and $s$ we define the $MFT$ operation, implemented in hardware as an element-wise optical multiplication followed by an \ac{OFT}:
        
    \begin{equation}
    \label{eqn:mft_operation}
        \begin{aligned}
            &MFT: (x, s) \mapsto z \quad x, s, z \in \mathbb{Z}^{n \times n} \\
            &MFT(x, s) = \mathcal{F}(x \cdot s) = z
        \end{aligned}
    \end{equation}
    
    Using these two identities, we can calculate the convolution of any two 2D arrays as follows:
        
    \begin{equation}
    \label{eqn:mft_operation_to_convolution}
        \begin{aligned}
            &I(MFT(MFT(x, 1), \mathcal{F}(y))) = \lambda (x * y) \quad \lambda \in \mathcal{R},
        \end{aligned}
    \end{equation}
    
    where $\mathcal{F}(y)$ can be derived using the $MFT$ function:
    
        \begin{equation}
    \label{eqn:mft_operation_to_convolution}
        \begin{aligned}
            &I(MFT(MFT(x, 1), MFT(y, 1))) = \lambda (x * y) \quad \lambda \in \mathcal{R},
        \end{aligned}
    \end{equation}
    
    where the ``inversion'' operator $I$ is a simple spatial reordering of the pixels. 
    (The need to include $I$ and $\lambda$ come from the outermost $MFT$ performing a direct \ac{FT} instead of an inverse one. 
    In practice, this is not a problem as $I$ has a linear complexity and can be fully parallelised, achieving a runtime in $\mathcal{O}(1)$.)
    	
    Unlike any other implementation of a convolution (e.g. decomposition into a matrix vector product, or \ac{FFT} followed by multiplication), this method does not suffer from the $\mathcal{O}(n\log{}n)$ or $\mathcal{O}(nm)$ penalty.~\footnote{The runtime of a fully parallelised \ac{FFT} may be smaller than $\mathcal{O}(n\log{}n)$, but will always be in $\Omega(\log n)$.} In this paradigm of computation the algorithmic complexity of the convolution function is $\mathcal{O}(1)$. As each output in a discrete convolution of two signals comprises a dot product, we may also use the convolution as a means to perform a dot product as in a dense layer of a neural network (including bias addition).
    	
	\section{Optical CNNs}
	
	\subsection{Architectures}
	    
	We created three \ac{CNN} architectures of varying width and depth, using max pooling and two dense layers with dropout. Architectures of varying width and depth were chosen to assess the effectiveness of the optical hardware for computer vision problems over a range of hyperparameters. A particular focus was to see if the networks increased in accuracy when larger numbers of filters were used. It was also desirable to evaluate the efficacy of a deeper network with more pooling layers, thereby focusing more of the problem into the convolutional layers (as opposed to relying more heavily on the dense layers in a shallower network). For these networks to be effective, meaningful hierarchical feature extraction via convolution layers must take place. There is also an argument that errors in the optical operations could reduce the performance of deeper networks. Weight updates rely on backpropagation which in our case assumes a conventional forward convolution, inviting a possibility for divergence between gradients (especially when propagated through multiple layers) and the true parameter update needed to minimise the loss function of the optical network. Training data was selected to be the MNIST dataset of handwritten digits \cite{lecun2010mnist} as it is in effect, a standard problem for assessing computer vision solutions.
	    
	\subsection{Calibration}
	    
	As the operations in the frequency domain necessitate complex control of the optical signal, calibration was required to control both the intensity and the relative phase of each pixel. First, `black' values were obtained for each pixel by finding configurations for the \ac{MZI} array where destructive interference was observed at the output of the grating coupler. After this a `white' value was found for a single reference pixel. This was achieved by choosing an arbitrary reference pixel, and controlling its \ac{MZI} element until a bright spot was observed at the grating coupler output. We selected a value giving an output slightly below the observed peak to account for variation in the brightness range of each pixel.
	    
	The next task was to find `white' values for each of the other pixels which were in phase relative to the reference pixel. We worked on each pixel independently. A pair of pixels that are both fully bright and in phase has a distinctive \ac{FT} pattern which was pre-calculated electronically. The \ac{MZI} for each pixel was then tuned until the best matching pattern was found as compared to the electronically calculated ground truth. A similar approach was taken for the values $-1, i$ and $-i$ which all have modulus equal to one but with phase offsets of $\pi, \pi/2$ and $3\pi/2$ respectively. Look up tables were generated for each of these values so that data could be translated into \ac{MZI} current. 
	    
    \subsection{Practical Considerations}
	    
	The thermal \ac{MZI} array induced some cross-talk such that certain combinations and interpolations from the calibrated look up tables were not successful in generating the desired input with a high enough degree of accuracy. For this reason, each \ac{FT} operation carried out on the device was done with each pixel independently, then reconstructed via electronic addition simulating a device with no thermal cross-talk. The prototype device used in this study had one \ac{MZI} per pixel. Using this configuration a convolution was achieved by taking the \ac{OFT} of the input and kernel functions, multiplying them in electronics, then taking the \ac{OFT} of the product.
	    
    While part of the calculation was performed electronically, we emphasise that the relevant data came from \ac{OFT} operations. In particular, sources of optical noise like residual ambient light, deviations from the Fresnel approximations, variations in laser power, and calibration errors are all included. The results shown below may thus be interpreted as showing the effects of optical errors on training a \ac{CNN} in a realistic experimental set-up using only off-the-shelf hardware components. Furthermore, such imperfections may actually have been exaggerated due to the multiple passes through the system required to calculate each \ac{OFT}. We further emphasise that this is a solution to problematic characteristics specific to the use of thermal modulation, and similar effects are not expected in systems that apply high speed components (such as ring modulators) that are standard in the field.
	    
	\section{Results}
	
\begin{figure*}
	\includegraphics[width=1.0\textwidth]{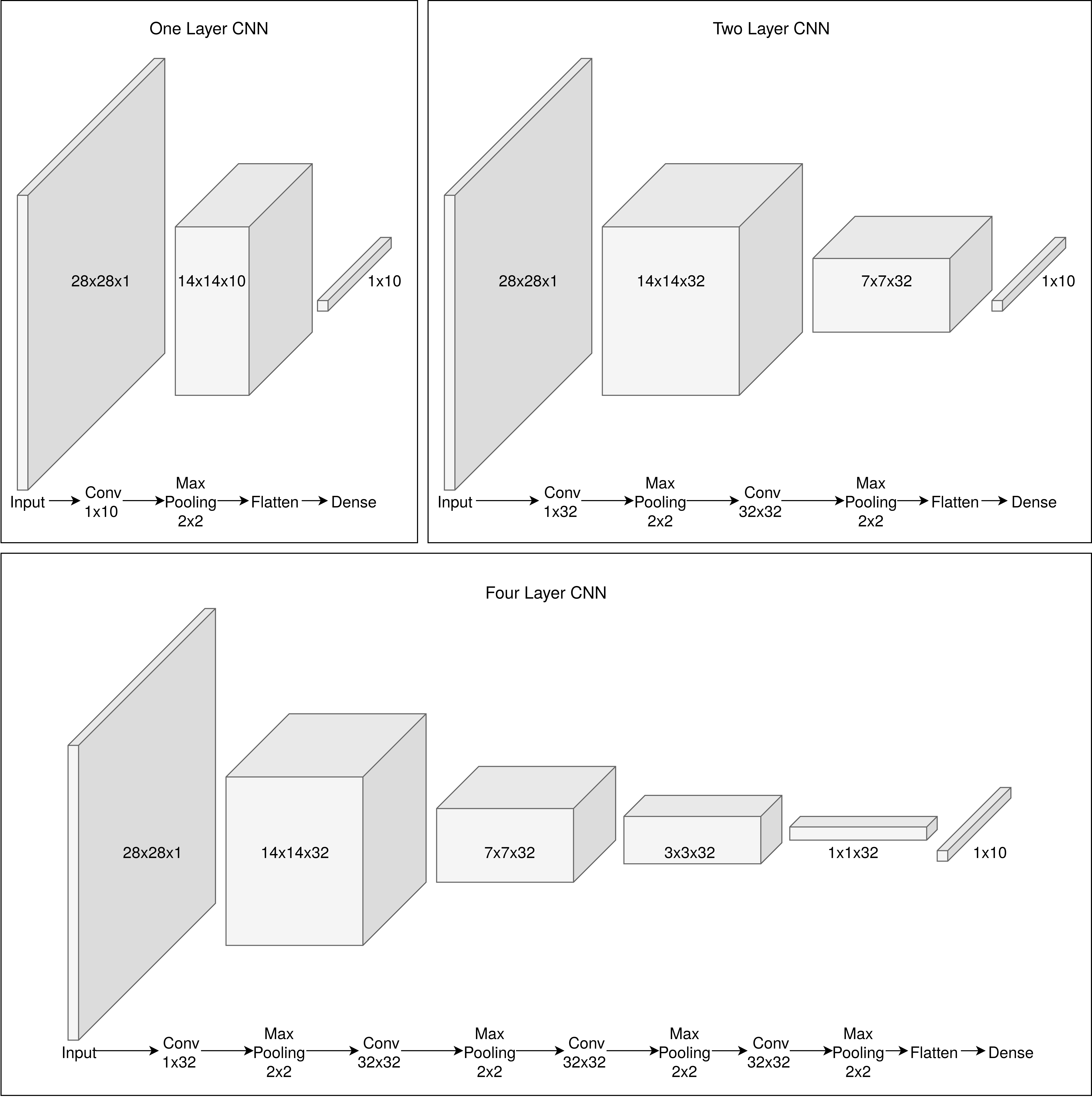}
	\captionof{figure}{Visual depiction of the \ac{CNN} architectures tested. Dimensionality of the 3D data in order height, width, channels.}
    \label{fig:cnn_architectures}
    \end{figure*}

	The \ac{CNN} was trained on the MNIST \cite{lecun2010mnist} dataset of handwritten digits. The optical \ac{CNN} was evaluated against an equivalent electronic network to evaluate the ability of the optical network to perform tasks in image classification with a conventional \ac{CNN} as a baseline. Three network architectures were trained (see Figure~\ref{fig:cnn_architectures}), each with a different number of convolutional filters and layers: 
	\begin{itemize}
	    \item{A single, 1x10-channel convolutional layer with 2x2 max pooling, followed by a pair of dense layers with dropout.}
	    \item{A 1x32-channel convolutional layer, followed by a 32x32-channel convolution layer, both followed by a 2x2 max pooling layer. Convolutional channels were followed by a pair of dense layers with dropout.}
	    \item{A network with four convolutional layers, each followed by 2x2 max pooling. The sizes of the layers 1, 2, 3 and 4 were 1x32, 32x32, 32x32, and 32x32 channels respectively. Due to the 28x28 input size, the output of the last channel was 1x1x32. This output was passed to a pair of dense layers with dropout.}
	\end{itemize}
	
	A summary of the results obtained is given in Table~\ref{tab:summary}:	
	\begin{center}
	\captionof{table}{Summary of results across the three networks, values shown are peak accuracy (\%) and minimum loss, electronic network results shown in parentheses.}
	\label{tab:summary}
	\resizebox{0.5\textwidth}{!}{%
        \begin{tabular}{ |c|c|c|c| } 
         \hline
         & One Layer & Two Layer & Four Layer \\ 
         \hline
         Train Acc. & $\mathbf{96.34}$ (97.24) & $\mathbf{97.13}$ (99.20) & $\mathbf{98.33}$ (99.40) \\ 
         Val. Acc. & $\mathbf{98.03}$ (98.52) & $\mathbf{98.50}$ (99.11) & $\mathbf{98.61}$ (99.35) \\ 
         Train Loss & $\mathbf{0.13}$ (0.11) & $\mathbf{0.10}$ (0.03) & $\mathbf{0.06}$ (0.02) \\ 
         Val. Loss & $\mathbf{0.06}$ (0.05) & $\mathbf{0.06}$ (0.03) & $\mathbf{0.04}$ (0.02) \\
         \hline
        \end{tabular}}
    \end{center}

    \begin{figure*}
	\includegraphics[width=1.0\textwidth]{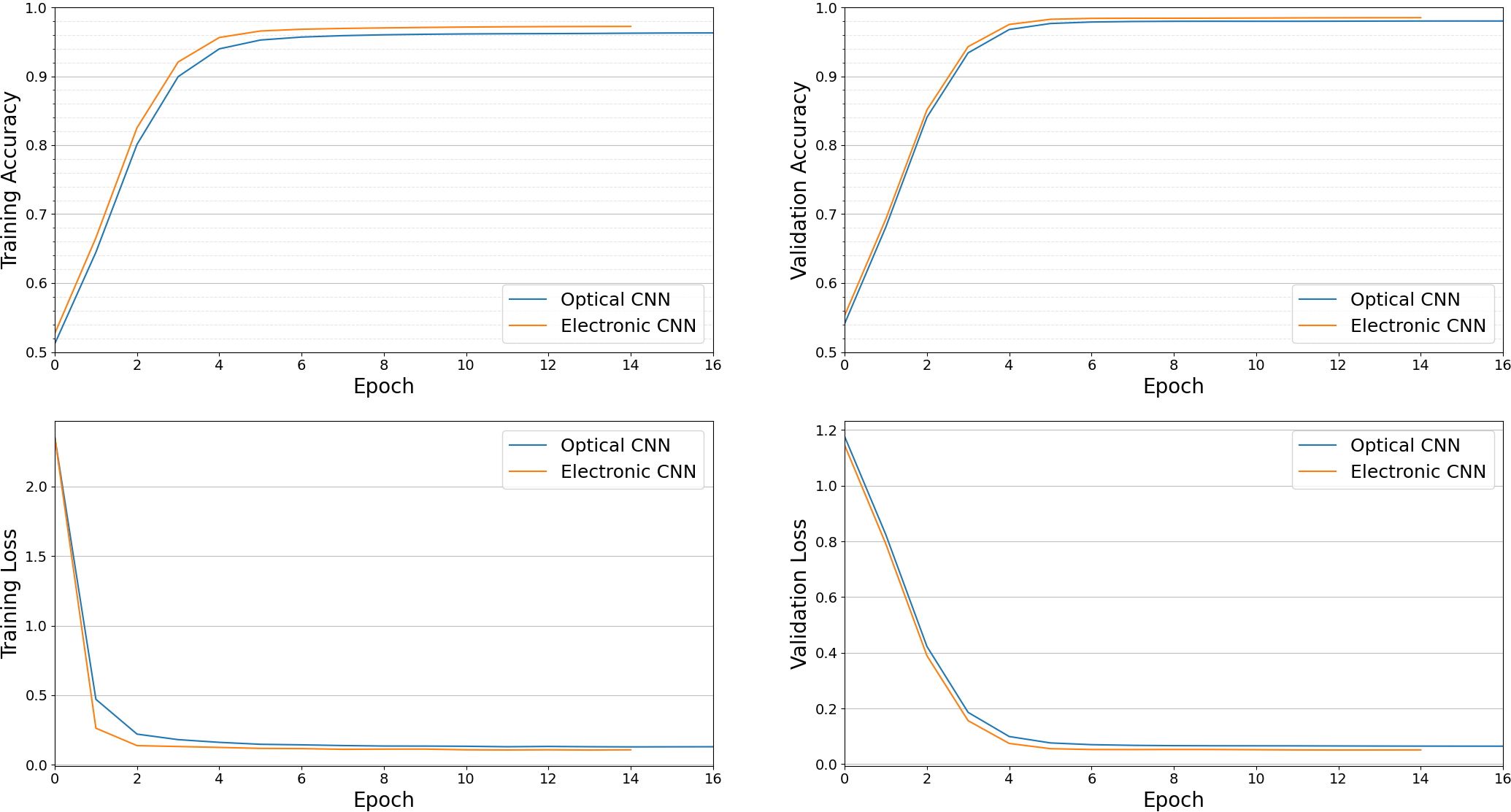}
	\captionof{figure}{One layer \ac{CNN} trained on MNIST. Training accuracy and loss (left). Validation accuracy and loss (right).}
    \label{fig:1l}
    \end{figure*}
    
    \begin{figure*}
	\includegraphics[width=1.0\textwidth]{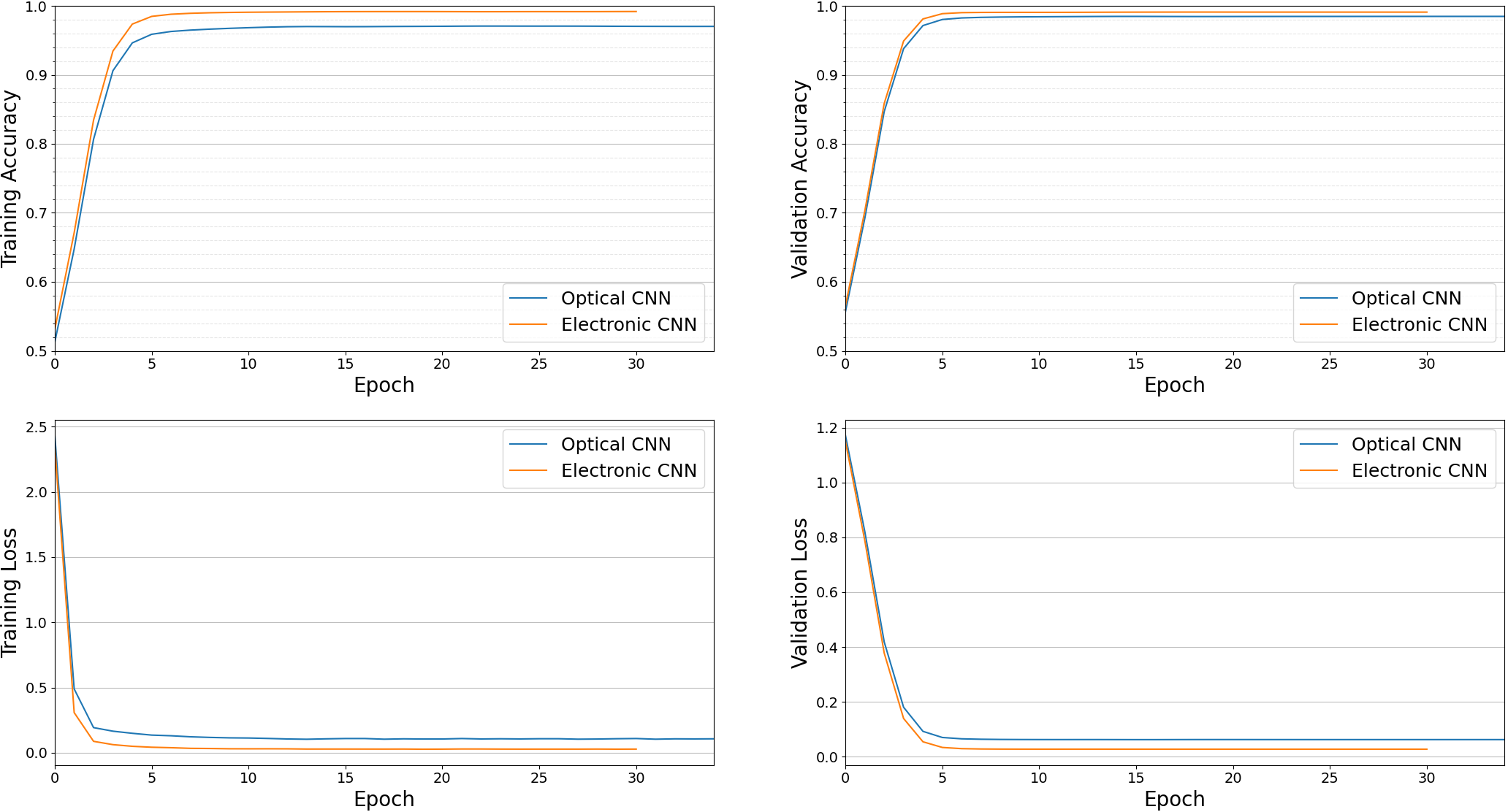}
	\captionof{figure}{Two layer \ac{CNN} trained on MNIST. Training accuracy and loss (left). Validation accuracy and loss (right).}
    \label{fig:2l}
    \end{figure*}
	   
	\begin{figure*}
	\includegraphics[width=1.0\textwidth]{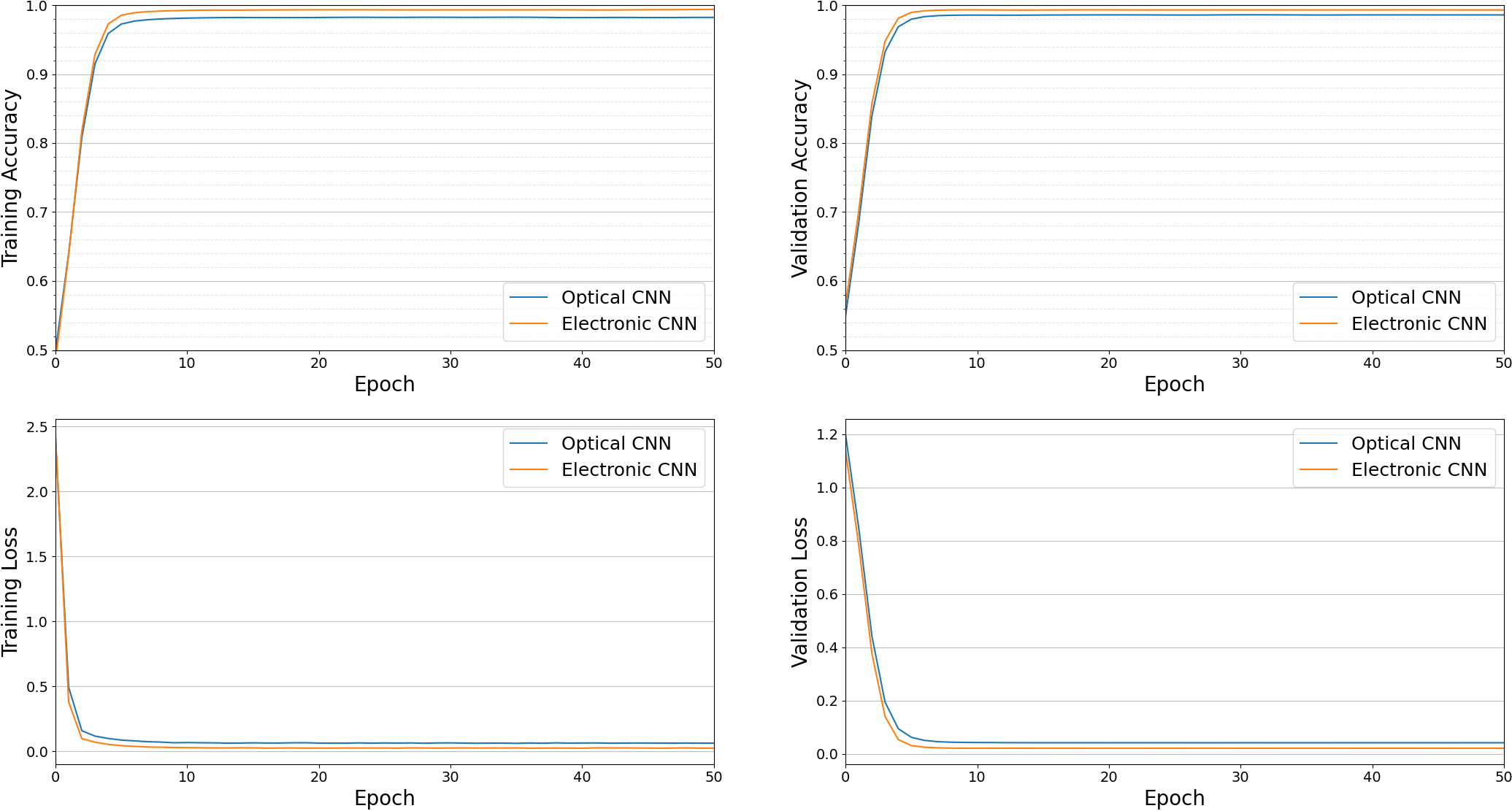}
	\captionof{figure}{Four layer \ac{CNN} trained on MNIST. Training accuracy and loss (left). Validation accuracy and loss (right).}
    \label{fig:4l}
    \end{figure*}
    

    As can be seen in the summary table (Table~\ref{tab:summary}) and training graphs (Figures~\ref{fig:1l} to~\ref{fig:4l}), the optical network consistently performed in a manner that is comparable to the electronic network. The electronic network did perform better overall, however the difference was always less than 1\% on validation accuracy metric. The optical network generally reached within 0.5-1\% of the validation accuracy observed in the electronic network. The optical networks also scaled well, with performance improving as the network grew larger in depth and width. This is encouraging as it supports the notion that deep \ac{CNN} architectures can be accelerated using optics without the divergence between forward and backward passes constituting a problem for training using gradients propagated through multiple layers. In the four layer network, the input was pooled down to the size of a single pixel. The accuracy attained by this network therefore indicates that performant hierarchical feature maps were constructed via optical convolutions.

    \subsection{Discussion}
    
    The optical \ac{CNN} performed well across all metrics and trained in a predictable way, reminiscent of the electronic equivalent and despite the limitations of the current heater-based prototype. However, these drawbacks were observed to be more limiting on training metrics, and not as severe on validation metrics, indicating that the network had learned patterns that generalise well to unseen data. \par
    
    While further experiments are needed to ascertain the origin of this behaviour, we conjecture that optical network will typically perform better on more challenging tasks, which would make them strong candidates for real-world applications in self-driving vehicles, medical image segmentation, or drone technology. Our argument is the following. Typically, even shallow neural networks can reach high test accuracy very fast on `simple' problems like MNIST because they learn the mapping between each input image and its label. This behaviour can lead to the under-specification problem \cite{d2020underspecification}: poor generalisability to unseen data that is drawn from different distributions to the training set. This problem can be partially solved by increasing the diversity of the training data, e.g. by using more samples, data augmentation~\cite{Shorten2019ASO}, or regularisation methods such as dropout~\cite{Hertz91,Hinton12}. \par
    
    Collectively, these techniques amount to introducing a form of randomness in either the inputs or how they are processed by each layer, improving the model's generalisability at the cost of slower training. As discussed, the optical network already has inherent sources of noise: fluctuations of the laser power, deviations from the optical approximations, ambient light, and imperfections in the electronic and photonic components invariably introduce small deviations from an exact convolution, some of which vary constantly. This prevents the model from learning an exact mapping between input image and label, thus reducing the ability of the model to overfit.
    
    We conjecture that optical noise may reduce overfitting and potentially increase generalisability without the need to introduce explicit regularisation. One could imagine, for instance, varying the input laser power to tune the amount of noise and find the optimal balance between speed of training and generalisability for each problem. Given the huge theoretical bandwidth, coupled with low latency and low power consumption of optical devices, this approach has the potential to provide previously unreachable levels of performance for artificial intelligence systems, and other applications where canonical methods are limited by the $log(n)$ scaling of the \ac{FFT} operation.
	 
	\section{Future Work}
	    
	Development work is underway to produce high-speed chip-level systems, based on existing \ac{SiPh} components and manufacturing processes. This will provide beyond state-of-the-art performance in convolutions, multinomial products, and other such operations that can be accelerated via spectral methods. Optalysys are also developing methods to apply the technology towards other high impact areas where these operations are central, including fully homomorphic encryption, partial differential equation solving, protein folding and post quantum cryptography. 

    \appendix
    
    \section*{List of Abbreviations}

    \begin{acronym}[CNNs] 
    \hypersetup{hidelinks} 
    \acro{CNN}{Convolutional Neural Network}
    \acro{FFT}{Fast Fourier Transform}
    \acro{FT}{Fourier Transform}
    \acro{SiPh}{Silicon Photonics}
    \acro{MZI}{Mach–Zehnder Interferometer}
    \acro{SLM}{Spatial Light Modulator}
    \acro{MFT}{Multiply \& Fourier Transform}
    \acro{LCD}{Liquid-Crystal Display}
    \acro{OFT}{Optical Fourier Transform}
    \end{acronym}
 
        \bibliographystyle{unsrt}
        \bibliography{library}
    
\end{document}